# Better support for collaborations preparing for large-scale projects: the case study of the LSST Science Collaborations
## Astro2020 APC White Paper

**Thematic Areas:** State of the Profession Considerations

**Principal Author:**
Name: **Federica B. Bianco**
Institution: University of Delaware,
 LSST SCs Coordinator,
 LSST Transients & Variable Stars SC co-chair
Email: fbianco@udel.edu

**Co-authors:**

  Manda Banerji (University of Cambridge, LSST Galaxies SC co-chair)
  John Bochanski (Rider University, LSST Stars, Milky Way, and Local Volume SC co-chair)
  William N. Brandt (Penn State University, LSST AGN SC chair)
  Patricia Burchat (Stanford University, LSST Dark Energy SC Deputy Spokesperson)
  John Gizis (University of Delaware, LSST Stars, Milky Way, and Local Volume SC co-chair)
  Zeljko Ivezić (LSST Project Scientist)
  Charles Keaton (Rutgers University, outgoing LSST Strong Lensing SC co-chair)
  Sugata Kaviraj (University of Hertfordshire, LSST Galaxies SC co-chair)
  Tom Loredo (Cornell University, LSST Informatics and Statistics co-chair)
  Rachel Mandelbaum (Carnegie Mellon University, LSST Dark Energy SC Spokesperson)
  Phil Marshall (SLAC, LSST, Past Dark Energy SC Spokesperson)
  Peregrine McGehee (College of the Canyons, LSST Stars, Milky Way, and Local Volume SC co-chair)
  Chad Schafer (Carnegie Mellon University, LSST Informatics and Statistics SC co-chair)
  Megan E. Schwamb (Gemini Observatory, LSST Solar System SC co-chair)
  Jennifer L Sokoloski (LSST Corporation Director for Science)
  Michael A. Strauss (Princeton, LSST Science Advisory Committee)
  Rachel Street (Las Cumbres Observatory, LSST Transients & Variable Stars SC co-chair)
  David Trilling (Northern Arizona University, LSST Solar System SC co-chair)
  Aprajita Verma (University of Oxford, UK, LSST Strong Lensing SC co-chair)

**Endorsers:** The LSST Project.

**Abstract**:

Through the lens of the LSST Science Collaborations' experience, this paper advocates for new and improved ways to fund large, complex collaborations at the interface of data science and astrophysics as they work in preparation for and on peta-scale, complex surveys, of which LSST is a prime example. We advocate for the establishment of programs to support both research and infrastructure development that enables innovative collaborative research on such scales.

## Key Issue and Overview of Impact on the Field

Many pressing, open questions in astrophysics, including the reason for the accelerated expansion of the Universe, the nature of dark matter and dark energy, the structure and evolutionary history of our own Galaxy and of the Solar System, and the dynamics of the ever-changing transient and variable sky, can be addressed with a wide-field, high-resolution, deep, rapid imaging survey of the sky. While targeted forerunner surveys are underway to investigate some of these problems (HSC [1], DES [2], KIDS [3], Pan-STARRS [4], ZTF [5], *etc.*), LSST [6] is an ambitious project that aspires to address *all of these questions and more* with an order of magnitude --at least-- increase in data rates, in number of transient discovered in real-time, and in number of objects for which information on position, velocity, brightness, color, morphology, and variability will be generated. **LSST represents a clear technical innovation, but that is not all: it is also driving a fundamental change in the functioning of the scientific community.**

Unlike previous surveys, **LSST does not have a science team tasked with generating *science* from its data**. The LSST Project designed the overall survey and is currently constructing the telescope and the pipeline to generate the original data catalogs and alerts from the LSST images. But it is the scientific community that has the responsibility of delivering scientific results that fulfill the tremendous potential of this project. **LSST will generate a revolutionary dataset, but the responsibility, privilege, and burden of turning it into science belongs to the public, and the scientific community in the US will have *unrestricted* access to the LSST data.**

Embracing this exciting challenge and the opportunities generated by the LSST survey, the scientific community has organized into *Science Collaborations* (SCs). **The LSST SCs are a unique, diverse, geographically distributed network of scientists collaboratively addressing questions ranging from fundamental physics to data science.** It is unprecedented that such a large swath of the scientific community would be actively working on a *yet-to-deploy* survey with a return on their work on a decade+ time frame, without receiving funding support from the project itself -- a survey that will ultimately produce data that will be accessible to the entire US (and Chilean) communities with no *preferred access for the scientists that are currently active*. The enthusiasm that the forthcoming LSST dataset has generated has indeed inspired over 1000 scientists to collaborate on science readiness and scoping, a testimony to the revolutionary potential of LSST!

**However, operating with minimal and unevenly distributed financial support has created significant challenges for the LSST Science Collaborations, which in turn compromises their planning and preparatory work for the Project, leading to potential vulnerabilities. This is the focus of this paper.**

The SCs have been and continue to be immensely valuable to the LSST Project. They provide scientific expertise that guides the survey design, including construction of data-processing pipelines [7, 8], plans for the survey strategy [9, 10], and insight that guides advocacy for the survey, educating other scientists and the public about the promise of LSST. **The SCs have done so for a decade, operating largely with no funding for **these activities[1].** As a chief example, the SCs are best placed to provide advice on the most scientifically productive survey observing strategies, and were specifically asked to do so [10]. But the unfunded collaborations struggle to enable contributions, as properly answering this question demands investigative work, software development, simulations, workshops, etc. Incomplete and/or inadequate answers however, compromise the project's overall scientific yield.

**The SCs represent a case study:** through the lens of our experience we can identify weaknesses in the US approach to funding ground-based science and advocate for a more effective structure of support for large, collaborative teams that tackle extremely ambitious, long-term science projects. In this paper we highlight the trend in astronomy that has now for many years seen ever-larger collaborations forming around projects of increasing complexity and requiring teams with increasingly diverse expertise —of which LSST is an exemplary case. In the following sections we describe the structure of the LSST SCs network, and the role and value of the SCs with respect to the LSST Project. We will articulate in §2 how the lack of funding has exposed the following risks: (1) duplication of effort, (2) inefficient use of resources, (3) (potential) ethical risks related to proper acknowledgment of work and effort, (4) (potential) insufficient effort to achieve maximal inclusion, and (5) (potential) loss of scientific discovery. We advocate for the creation of clear paths to fund: (1) scientific scoping by large collaborations, (2) scientific preparation for complex projects with long lead times, (3) development of software to support scientific discovery to minimize duplication of effort across team, (4) development of infrastructure and protocols for early and effective collaboration, to maximize the scientific productivity of the LSST and future projects of similar scale, and to do so ethically and inclusively.

## 1. The role and structure of the SCs

The SCs were originally formed by the LSST Project in 2008 to provide a forum to engage the community in interacting with the LSST Project, and to make the scientific case to be presented to the 2010 Decadal Survey [11]. But they are today independent of the LSST Project: eight teams, self-governed and self-managed, that gather over 1000 scientists from six continents (Figures 1 & 3). The original teams have evolved since 2008: some teams dispersed, others merged, and some new teams emerged, reflecting the changing landscape of astronomy research in the past 10 years. For example, the Informatics and Statistics SC was created in late 2009, testifying to the rising importance of data science and machine learning in astronomy. The current breakdown and membership is shown in Figure 1.

Composed of astronomers, astrophysicists, and data scientists, the SCs are a diverse pool of experts that study radically different phenomena in the realms of natural science and data science, ranging from Solar System studies of objects only tens of thousands of miles from the Earth, to the study of objects at distances so great that they can only be measured in time-past, all the way to the visible edge of the Universe. **The** comprehensive heterogeneous science portfolio that LSST will enable requires a broad set of expertise, and an exploration of the physical Universe of this scale and complexity would not be possible within a single team: the purpose of the LSST SCs is to bring a diverse pool of scientists together to make the whole greater than the sum of its parts.

---

[1] DESC is partially supported by the Department of Energy DE-AC02-05CH11231. We discuss this, including the consequences of funding imbalances within the SCs, in §2.1 and 2.3.

## 1.1 Current Structure of the SCs

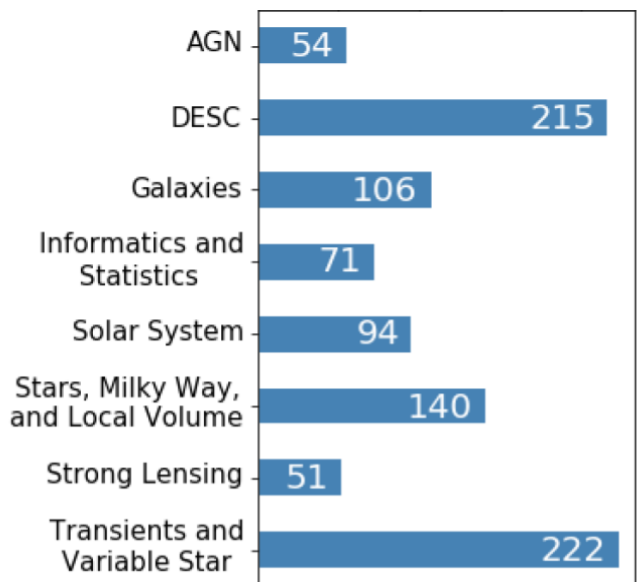

Figure 1: LSST SCs breakdown and membership. For DESC the membership indicated only includes "full members"; the total DESC membership exceeds 900. For all other SCs, even if they have a two-tier membership structure, like TVS, the entire membership count is shown.

**The LSST SCs are a complex network**: each SC is a node in the network (Fig. 2) with each node identified by its members' scientific interests and expertise. While each SC pursues specific scientific goals, these investigations benefit from a synergistic approach reaching across SC boundaries. For example, the study of our Galaxy advances through both studies of resolved stellar objects, largely the focus of the Stars, Milky Way, and Local Volume (SMWLV) SC, as well as from the stars' variability characteristics, a focus of the Transients and Variable Stars (TVS) SC. Individual scientists can participate in more than one SC. This fosters intercommunication between SCs. *Each node is developing SC-specific expertise and software, but the umbrella structure of the SCs is in place to allow them to integrate.* The members of the different SCs have subject-matter overlaps but also, critically, technical overlaps and infrastructural improvements would have huge impact where the same technical problem must be solved for multiple science cases, and such overlap is especially common for data science technology, *but the infrastructure of the SCs and its development are minimally supported.*

**Internally, most collaborations are also subdivided into working groups, subgroups, or task forces.** The most striking example is the TVS SC, which studies all phenomena of the transient and variable sky: anything that, contrary to popular intuition about the sky being immutable, changes in luminosity, color, position, or shape, from exploding stars to planet transits and microlensing. Reflecting the breadth of this science, the TVS SC is subdivided into 15 subgroups focusing on different phenomena, and several annual "Task Forces" are created to address urgent needs that straddle more than one subgroup. The diverse internal structure of the SCs, however, affects their effectiveness, as we will discuss in (§2.3)

## 2. Vulnerabilities arising from or exacerbated by the lack of funding

We identify the following primary vulnerabilities in the current LSST SC network that arise from lack of funding to support this large collaboration.

**1. The need for intense collaborative research:** high-quality LSST science can only succeed in a timely manner if a significant number of SC members can access funds that would allow them to simultaneously, collaboratively, work on shared research challenges (§2.1, 2.2 and 2.3).

**2. Funding disparities:** Every US-affiliated scientist inherits data rights, but access alone is not sufficient to produce science. US SC members have no direct funding for LSST-based research (except for some DOE-supported tasks for DESC)**,** while many international counterparts plan their investments in LSST including support for science exploitation. This creates the specific risk that *US scientists will be unable to compete with funded international LSST members and lead LSST-based science.* Support for science and for the development of a solid infrastructure and rules of engagement would level the playing field across the SCs (§2.3, 2.4)

**3. Unstructured communication.** This creates the risk of inefficiencies due to redundant efforts, and it complicates the refinement of the LSST survey, which is pulled in directions that are at times orthogonal when better communication could realign priorities. The resulting inefficiencies may impair or even prevent the pursuit of specific science goals by putting unnecessary stress on the limited computational resources available to process such a complex dataset. The lack of effective communication is largely a consequence of the lack of support for the infrastructure of the SCs (§2.3).

**4. Inhomogeneity in the level of organization of the SCs:** This affects the ability to effectively communicate with, between, and within the SCs, and to optimally distribute resources and funds. It heightens the risk of conflicts between nodes (§2.3).

**5. Geographical segregation and the need for interaction:** Each node of the network is geographically distributed. While this is a desirable feature of the network, in-person interaction on a regular basis is necessary to align the work and goals of each node. However, such interaction is costly and funds are not generally available to support frequent meetings (§2.4).

## 2.1 Current funding and consequences of lack of support.

Here we describe the existing funding streams available for the SCs and highlight their inadequacy in supporting research and infrastructure development. With the exception of the Dark Energy Science Collaboration (DESC), the SCs are not funded or supported as entities by US-agency funds.

The individual members of the SCs can apply, and have applied, for funding through traditional mechanisms such as NSF Astronomy and Astrophysics Research Grant (AAG) solicitations and grants from private foundations. However, the SCs are carrying out LSST preparatory work with LSST data many years to come (a lead time of 14 years when the SCs were created!). This has meant that many proposals for funding cannot promise prompt *science* deliverables, and are therefore at a disadvantage in competing with non-LSST proposals that focus on existing data and short-term science results. While one could imagine testing software under development for LSST on precursor surveys, in many cases precursor surveys lack the essential characteristics (depth, cadence, area, etc.) that define the LSST dataset, or are not publicly available; hence this is not necessarily a viable option. More importantly, **an exploration of the physical Universe of the breadth, scale, and complexity that LSST aspires to requires** **diverse expertise combined in a coordinated effort, rather than fragmented work done in small, independent teams competing for funding to address narrow science targets. A collaborative effort at this scale requires support for collaborative research and support for infrastructure to facilitate communication and collaboration.** The high-energy physics community, accustomed to large collaborations, has a more effective way to support the science community in the lead-up to project operations. DOE is supporting some of the infrastructure of the DESC in the same fashion in which high-energy physics projects are traditionally supported. We will address issues related to this funding imbalance in §2.3. Similarly, space missions are generally effectively supported by grants that cover the building of the hardware, software, and pipeline from data reduction to production of the target science deliverables of the mission. Meanwhile, **the LSST, a novel ground-based astronomy project of unprecedented scale, is highlighting a gap in the US science funding model (§2.2).**

The *LSST Corporation [12]*—a 503(c)(3) non-profit organization charged with obtaining philanthropic support for science (and formerly operations) leading up to and during LSST— has provided the SCs with some valuable infrastructure support[2]. In the last five years they distributed ~$500K to the SCs through the LSST Corporation Enabling Science program, a call open to the entire community but to which the SCs have consistently applied and obtained support for meetings, undergraduate research internships, and some science projects [13]. *The Enabling Science funds have now been exhausted.* Although the Corporation is reaffirming its commitment to fundraise for and support the SCs and LSST-related work, they estimate that *the need for support of LSST science support exceeds by one-to-two orders of magnitude what they have been providing.*

**DESC is supported as a collaboration by the Department of Energy (**DOE, Contract no. DE-AC02-05CH11231) **for investigating the nature of Dark Energy**; prior to this contract, limited research support was available to DESC members at DOE labs and to University-based PIs through the Cosmic Frontier competitive grant system. The support for DESC has enabled the scoping of dark-energy related science for LSST including developing collaboration infrastructure, producing, processing, and distributing data simulations, and analysis software that works at the needed scale and precision [*e.g.* 8, 14]. Support for DESC's infrastructure is crucial to enable dark energy science with LSST data, and other science collaborations would benefit from a similar steady source of infrastructure support to enable their work towards their science goals (§2.3).

## 2.2 The direct impact of missing research funds on science productivity.

**To truly fulfill the promise of LSST, and obtain the maximum return for the investment in LSST that the NSF and DOE made by supporting LSST's construction, this large and diverse community of scientists needs support to work collaboratively.** In the current model, the vast majority of the LSST SC members

---

[2] LSSTC has supported the SCs with access to communication tools (Slack, Bluejeans tele-conferencing software), hosting web pages and membership databases, and with support for the LSST Science Collaborations Coordinator equivalent to one month of summer salary and supporting SC related travel.

(with the exception of some DESC and international members who are funded to different degrees for their SC-related activities) *participate in LSST-related work in their "free" or "independent research" time that is supported at a low level by their regular jobs.* That is enough for some of the members to participate in telecoms and occasional meetings, but not enough, in general, to engage in innovative work; enough occasionally to generate research products in small teams, but not to coordinate research efforts and leverage the diverse expertise in the network to produce truly innovative science!

## 2.3 The impact of lack of funding and funding imbalances.

Without support, the SCs cannot devote the necessary effort to organize appropriate communication and run the risk of producing redundant, duplicated effort, inefficiently pursuing tasks that are similar, without unifying their goals. This inefficient effort, as measured in human and computational resources, will **impair the investigation of more science goals**, and, without a unified plan, **will attempt to push LSST's strategy in different, incompatible directions, reducing the overall science throughput.**

The non-homogeneous structure of the SCs (§1.1) is a reflection of the "grassroots" nature of the network, but it is not without consequences. As of today the SCs have different levels of organization. Activities like the creation of charters and publication policies require time, dedication, and research (both library research and survey research within the collaboration). DESC's management structure was created as a part of DESC's responsibilities toward the DOE, which provides funding for their operations. Due to lack of funding to support the administrative activities of the SCs, the definition of a governance structure and creation of supporting documentation have progressed at different pace and generally slowly within the SCs. Without the definition of roles and responsibilities it is unclear who to refer to when soliciting insight from the SCs or requesting the development of science or software, and without support to undertake those responsibilities, it is difficult for the SCs to deliver what was requested with consistently high quality and in a timely way.

**An important consequence of the funding imbalance between DESC and the other SCs is that the SCs are not all in the same position to advocate for their science priorities: a better organization enables better communication and advocacy.** Consider, for example, the November 2019 *Call for Cadence White Papers* issued by the LSST Project to finalize the LSST observing strategy [10]: while the call was open to the entire scientific community, the vast majority of authors were affiliated with one or more SCs. The breakdown of submissions in response to this call is shown in Fig. 5. Almost all SCs participated by leading the submission of one or more white papers. However, the SCs with support for research and established managerial roles and communication channels can better strategize to (1) generate a coordinated response and (2) support their science case with quantifiable metrics. DESC was able to produce a single, coordinated response to the call (a response for each survey within LSST) and produce metrics to evaluate each science driver of their cadence proposals. Meanwhile the TVS SC, for example, responded with as many as 20 TVS-lead papers, demonstrating the enthusiasm and dedication of the members, but could not undertake a large, coordinated effort to merging these science cases into fewer, stronger proposals as originally envisioned. *Thus, some science cases may be weakened by this dispersion*

*in spite of their inherent value, leading to a suboptimal strategy design for LSST.*

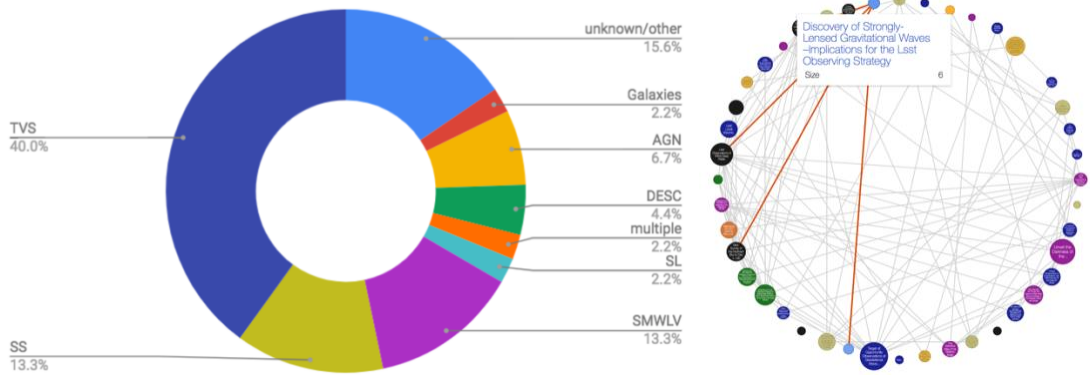

Figure 2: The breakdown of 2019 Cadence White Paper (WP) submissions by the Science Collaboration of the leading author (left) and the coauthorship network (right). An interactive version of this plot is available at http://fbb.space//LSSTWP/LSSTwhitePapers.html. The left panel shows the fraction of WPs submitted by a SC according to the membership of the paper's first author. On the right side, each circle represents a paper, the size of the circle reflecting the size of the authoring team, color-coded as in the left panel by lead author SC, and the papers are linked by their co-authors. Only 15% of the papers were submitted by first authors unaffiliated with a SC.

Furthermore, as indicated by the LSST SAC review of the white papers [15] *most proposers were not able to develop metrics codes to support their proposals.* Creating and coding metrics requires sustained effort to connect low-level quantities produced in simulations of the LSST survey to the high-level observables that determine scientific success, effort to study the LSST simulation outputs [16], and to code metrics within the LSST MAF API [17], all of which is time consuming. We attribute the general failure to deliver metrics to the lack of support for members of the SCs to pursue SC-related research activities. More generally, DESC's productivity will remain unmatched by other SCs without support for all SC's research and operations, regardless of the inherent importance of the science they each pursue.

### 2.4 International participation.

Several countries outside of the US have invested in LSST science, drafting agreements (now under revision) to acquire data rights, and supporting science development and preparatory work within their community: as a result, a significant fraction (about 1/3) of the membership of the SCs are from countries other than the US (Fig. 3). In cases where international members are supported by their funding agencies for their LSST work, rather than engaging in it as an extra-curricular activity or an activity that can be performed in a small fraction of "independent research" time (see §2.2) their productivity can exceed that of US members. For example, while the US dominates membership in the SCs, in the November 2019 *Call for White Papers* [10]*,* 46% of the submitting authors were not affiliated with US institutes. This is despite the fact that the US community had a many-year lead ahead of most other countries. **If the US scientists are to lead the discovery with this US-financed project, the US scientific community needs immediate support to prepare for LSST.**

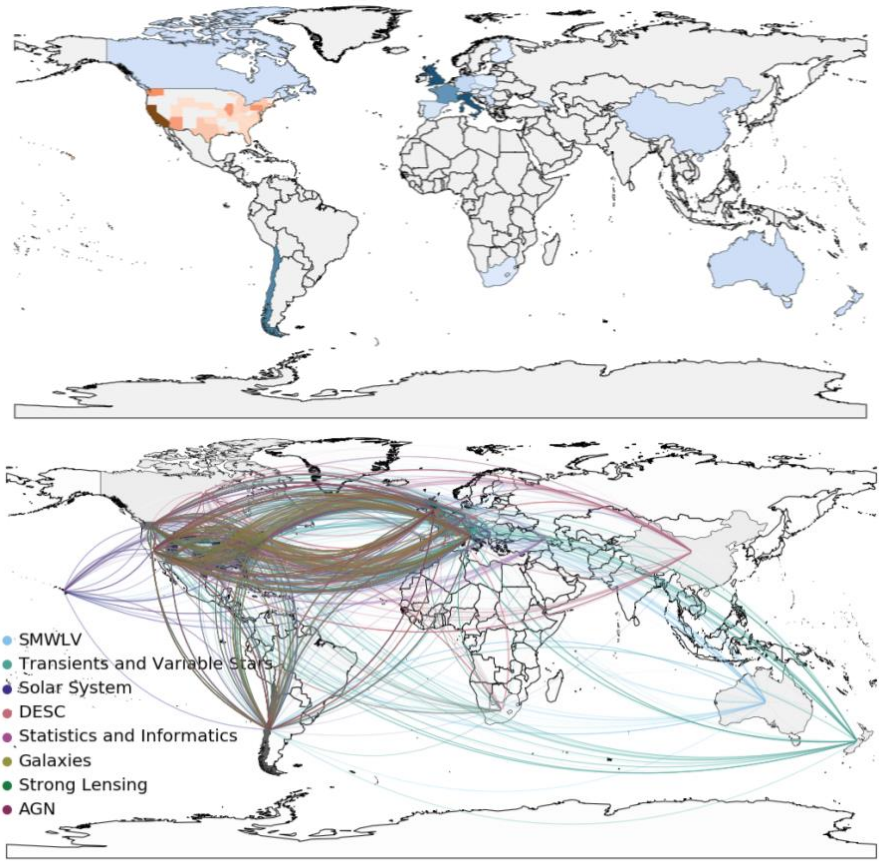

Figure 3: *Top*: choropleth of the LSST SC membership. The color intensity is proportional to the number of members. The US is divided by States, with membership indicated by the intensity of orange, all other countries' membership is indicated by the intensity of blue. *Bottom*: the geographical network structure of the LSST is shown with links connecting members of the same SC across the world.

## 2.5 Inclusion and diversity.

It is now fully acknowledged, although our community still has a long way to go, that inclusion is an important issue in the sciences and that maintaining an inclusive diverse community in any STEM effort is a matter of social justice, as well as a way to foster creativity and excellence [18]. Fostering inclusion requires deliberate effort, time to self-educate and strategizing, and specific activities leading to inclusion [19, 20]. Without funds to support them, the SCs are unable to devote sufficient effort to these activities. *Currently, the SCs are in fact the most diverse element of the LSST ecosystem (e.g. the overall women representation is close to 30% and increasing over time, and 7 out of the 15 current chairs are women)* **but they are not out of the danger of falling back on a less progressive representation of women and minorities, particularly as we get closer to LSST operations and science programs begin productivity [21]. At this stage, there is a risk the atmosphere can turn competitive, and more aggressive -- an environment which is known to disadvantage women and minorities**, and early career scientists.

We note that the NSF has recently recognized the value of the educational opportunities that arise with a project as innovative as LSST, and joined the LSST Corporation and a number of philanthropists in supporting the LSSTC Data Science Fellowship (https://astrodatascience.org), which has the aim to train a new generation of data-wise scientists while fostering an inclusive and diverse community. However, this effort is not sufficient to fulfill the needs for a well-trained workforce, as many more graduate-level students as well as postdocs trained in data-driven inference on peta-bite scale surveys are needed to fulfill the scientific promise of LSST.

## 3. Specific recommendations

## Strategic Plan:

From our experience within the SCs, we recommend that more opportunities be provided to enable funding of large collaborative structures that tackle ambitious, long-term goals. Specifically, we recommend:

1. Rewarding funding proposals that address R&D questions or infrastructure needs identified as high priority for meeting the science goals of the SCs.
2. Rewarding proposals that include significant organizational and managerial work when in the context of enabling the functioning and coordination of large, diverse collaborations. This will allow large collaborations to equalize the contribution of each node of a large network, like the LSST SCs, enabling the merit of science cases to be assessed independently of availability of funds (some recent NSF calls, e.g. www.nsf.gov/pubs/2019/nsf19548/nsf19548.htm, may be suitable for this). It will allow the development and implementation of strategies to foster and support a diverse inclusive community. The development of technical documents and white papers, generally only recognized at the level of service in academic careers, should also be acknowledged and rewarded.
3. Creating more positions at the postgraduate and graduate level that facilitate working with multiple mentors. This may require enabling distribution of funding and responsibilities across different departments and institutes, for example as fellowships co-sponsored by multiple institutes or long-term scholarships. This will reduce duplication of effort and generate intersectional, interdisciplinary research products, as well as fostering collaborations between institutions.
4. Rewarding proposals that produce open and shared software tools (the focus of the Astro2020 white paper: Tollerud et al. 2019) and data products, disseminated via professional archive services to ensure longevity of public access.
5. Supporting proposals for interdisciplinary meetings and workshops, to ensure regular knowledge transfer between different scientific communities.

The LSST SCs have only 3-years lead time to begin operations, over which time they must: develop software to transform the LSST data products into science results, advance theoretical fields to generate predictions that can be tested with the LSST data, collect datasets that can be used in preparation and in conjunction with the LSST data, and plan and secure alignment of follow-up capabilities to enable effective coordinated follow-up studies of LSST targets with worldwide facilities spanning the electromagnetic spectrum. Funding the LSST SCs as outlined above is a necessary and critical step to assure the scientific return of the current NSF and DOE investment in LSST.